\documentclass[prl,twocolumn,nofootinbib,superscriptaddress,showpacs]{revtex4-1}

\usepackage{color,amsthm,amsmath,amsfonts,graphicx,bm}

\usepackage{bm}
\usepackage{bm,eufrak}
\usepackage[export]{adjustbox}
\usepackage{amsfonts}
\usepackage{amssymb}
\usepackage{amsmath,mathrsfs}
\usepackage[pdftex]{hyperref}
\usepackage{epsfig}
\usepackage{graphicx}
\usepackage{enumerate}
\usepackage{multirow}
\usepackage{verbatim}
\usepackage{subfigure}
\usepackage{bbold}
\usepackage{soul}
\usepackage{scrextend}
\usepackage{tikz}
\usepackage{comment}
\allowdisplaybreaks
\usepackage{hyperref}
\usepackage{float}

\newcommand{\ket}[1]{\vert#1\rangle}
\newcommand{\bra}[1]{\langle#1\vert}
\newcommand{\tr}{\mathrm{tr}}

\newcommand{\mr}[1]{\mathrm{#1}}
\newcommand{\dg}{\dagger}

\renewcommand{\tilde}{\widetilde}

\newcommand{\Tr}{\operatorname{Tr}}

\newcounter{notes}
\stepcounter{notes}

\DeclareFontFamily{OT1}{pzc}{}
\DeclareFontShape{OT1}{pzc}{m}{it}{<-> s * [1.10] pzcmi7t}{}
\DeclareMathAlphabet{\mathpzc}{OT1}{pzc}{m}{it}

\def\tcm{T.C.M. Group, Cavendish Laboratory, University of Cambridge, J.J. Thomson Avenue, Cambridge, CB3 0HE, UK}
\def\DAMTP{DAMTP, University of Cambridge, Wilberforce Road, Cambridge, CB3 0WA, UK}

\begin{document}

\title{Local integrals of motion detection of localization-protected topological order}

\date{November 2021}

\author{Thorsten B. Wahl}
\affiliation{\DAMTP}
\author{Florian Venn}
\affiliation{\DAMTP}
\author{Benjamin B\'eri}
\affiliation{\DAMTP}
\affiliation{\tcm}

\begin{abstract}
Many-body-localized (MBL) phases can be topologically distinct, but distinguishing these phases using order parameters can be challenging. 
Here we show how topologically distinct local integrals of motion, variationally parametrized by quantum circuits, can be used to numerically demonstrate the topological inequivalence of MBL phases. 
We illustrate our approach on a fermionic chain where both topologically distinct MBL phases and  benchmark comparisons to order parameters are possible. 
We also use our approach, augmented by the DMRG-X algorithm, to extract high-energy topological doublets. 
We describe applying our methods to higher dimensions to identify MBL topological order and topological multiplets  hidden by the dense many-body spectrum. 
\end{abstract}

\maketitle

\section{Introduction}
 
Many-body localization (MBL)~\cite{Fleishman1980,gornyi2005interacting,basko2006metal,znidaric2008many,pal2010mb,Bardarson2012,imbrie2016many,NandkishoreHuse_review,AltmanReview,Abanin2017,Alet2017,ImbrieLIOMreview2017} has attracted a wealth of interest in the last fifteen years. 
One of the most striking features of MBL is the violation of the eigenstate thermalization hypothesis~\cite{deutsch1991quantum,srednicki1994chaos}: 
MBL systems do not thermalize, but instead retain some memory of their initial state. 
As a result, some MBL systems are able to protect quantum information~\cite{2013Bauer_Nayak,Huse2013LPQO,bahri2015localization,Goihl2019}.

MBL systems are characterized by local integrals of motion (LIOMs)~\cite{chandran2015constructing,Rademaker2016LIOM,ImbrieLIOMreview2017,Abi2017,Goihl2018}: exponentially localized operators commuting with the Hamiltonian and each other. 
As a result, all eigenstates of MBL systems obey the entanglement area law~\cite{Friesdorf2015}. 
Therefore, topological order~\cite{Kitaev2006,Levin_Wen2,topQC}, normally  present only in ground states, can also occur in high-energy MBL eigenstates~\cite{2013Bauer_Nayak,Huse2013LPQO,kjall2014many,bahri2015localization,2015Slagle,Thorsten,1DSPTMBL,2DSPTMBL,topMBL}. 
Nonetheless, owing to the lack of local order parameter, and due to the overlap (in energy) of topological multiplets away from the strongly MBL regime or beyond one-dimension (1D), numerically detecting  the topology of MBL phases remains challenging for reasons beyond the mere exponential scaling of the Hilbert space~\cite{Parameswaran2018}.

In this work, we show how a topological LIOM framework~\cite{topMBL} (with LIOMs and tLIOMS for topologically trivial and nontrivial cases, respectively), combined with quantum circuits for MBL~\cite{Wahl2017PRX},  can be used to numerically detect topological MBL.
Furthermore, as we also show, when used in conjunction with the excited-state density-matrix renormalization group (DMRG-X)~\cite{Khemani2016MPS,Yu2017}, this approach can also  identify topological multiplets provided the system is deep in a topological MBL phase.

Due to its polynomial scaling with system size, and the generality of tLIOMs, our approach is a general way to capture topological MBL, including beyond 1D. 
Nonetheless, to demonstrate its use, we focus on 1D: we study the disordered interacting Kitaev chain, displaying two topologically distinct MBL phases~\cite{Kitaev_chain,Huse2013LPQO}.  
Studying this system is useful not only due to the exact diagonalization (ED) benchmark available in 1D, but also due to a local order parameter benchmark available thanks to a duality to a system displaying conventional symmetry breaking.

\section{(Topological) LIOMs from quantum circuits}
 
LIOMs are typically assumed to be related to Pauli-$z$ operators $\sigma_i^z$ acting on site $i$ via a local unitary transformation $U$, $\tau_i^z = U \sigma_i^z U^\dg$.  
The $\tau_i^z$ are thus exponentially localized.
They define a complete set of quantum numbers since, $[H, \tau_i^z] = [\tau_i^z,\tau_j^z] = 0$ $\forall \, i,j = 1, \ldots, N$, with $N$ the system size. 
The locality of $U$ implies that the eigenstates of $H$ are local-unitary related to local product states: $H$ cannot display topological order~\cite{Bravyi2006,Chen_Gu,*HastingsPRL2011}.
Hence, for topological MBL systems the notion of LIOMs has to be extended~\cite{topMBL}: 
One must use tLIOMs, given by $\tau_i = U S_i U^\dg$, where $U$ is again a local unitary, but $\{S_i\}$ is now a set of mutually commuting local stabilizers~\cite{Gottesman1997thesis,NielsenChuang} 
whose common eigenstates all display (the same) topological order.  
On a topologically non-trivial manifold, to get a complete set of quantum numbers one augments tLIOMs by the non-local $\tau_i^\mathrm{nl} = U S^\mathrm{nl}_i U^\dg$, where $S^\mathrm{nl}_i$ are non-contractible Wilson loops (i.e., logical operators)~\cite{topMBL}.

To utilize tLIOMs numerically, we use that owing to its locality, we can efficiently approximate $U$ by a fixed-depth quantum circuit $U_\mr{SR}$~\cite{Wahl2017PRX}.
We employ $U_\mr{SR}$ variationally: we aim for $[H, \tau_i^z] = 0$ by minimizing
\begin{align}
f &= \frac{1}{2} \sum_{i=1}^N \tr \left([\tau_i^{},H^{}] [\tau_i^{},H^{}]^\dg\right), \label{eq:fom}
\end{align}
where $\tau_i^{} = U_\mr{SR} S_i U_\mr{SR}^{\dg}$ and $S_i$ are topological  stabilizers for a tLIOM ansatz, while $S_i = \sigma_i^z$ for a conventional LIOM ansatz.  
The lower $f$, the better the approximate (t)LIOMs describe the system. 
In the topological case, the non-local $\tau_i^\mathrm{nl}$ enter only for  multiplet splittings. 
Hence, they almost commute with $H$ if all tLIOMs are optimized and thus can be omitted in Eq.~\eqref{eq:fom}.
To minimize $f$, we proceed similarly to Ref.~\cite{Wahl2017PRX}: 
We expand $H$ as a sum of local terms and identify which combination of local terms contributes to $f$ (which is quadratic in $H$). 
Each of the contributions is then represented as a tensor network contraction, which we evaluate efficiently using common numerical methods~\cite{Wahl2017PRX,Smith2018,Lewenstein2020}.
While developed for non-topological systems, this approach directly applies also to the topological case: by using topological $S_i$ (which act on multiple sites) we merely increase the number of tensors in each contraction. 
In the non-topological phase, we expect the conventional LIOM ansatz to perform better. 
In the topological phase, however, the tLIOM ansatz will minimize $f$.

\section{Model} 

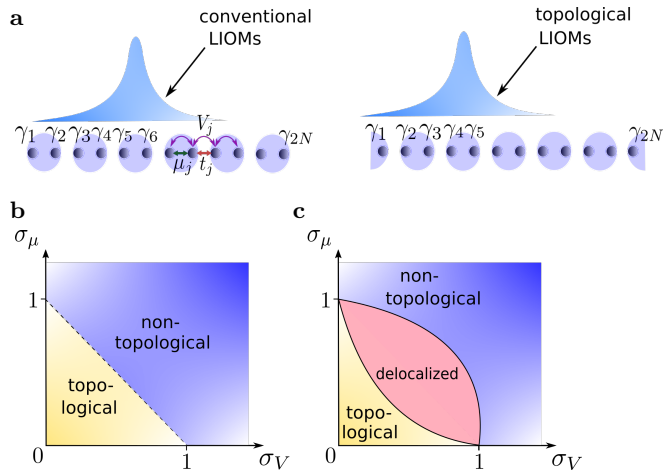
\begin{figure}
\begin{picture}(250,70)
\put(0,0){\includegraphics[width=0.48\textwidth]{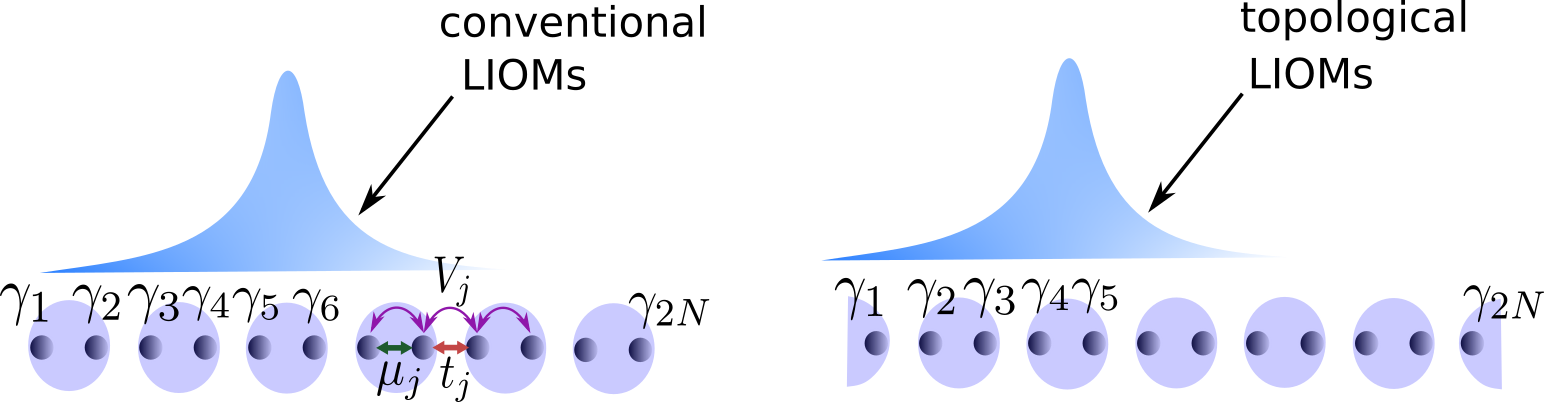}}
\put(-2,57){\textbf{a}}
\end{picture}
 \\
\begin{picture}(250,110)
\put(0,0){\includegraphics[width=0.42\textwidth]{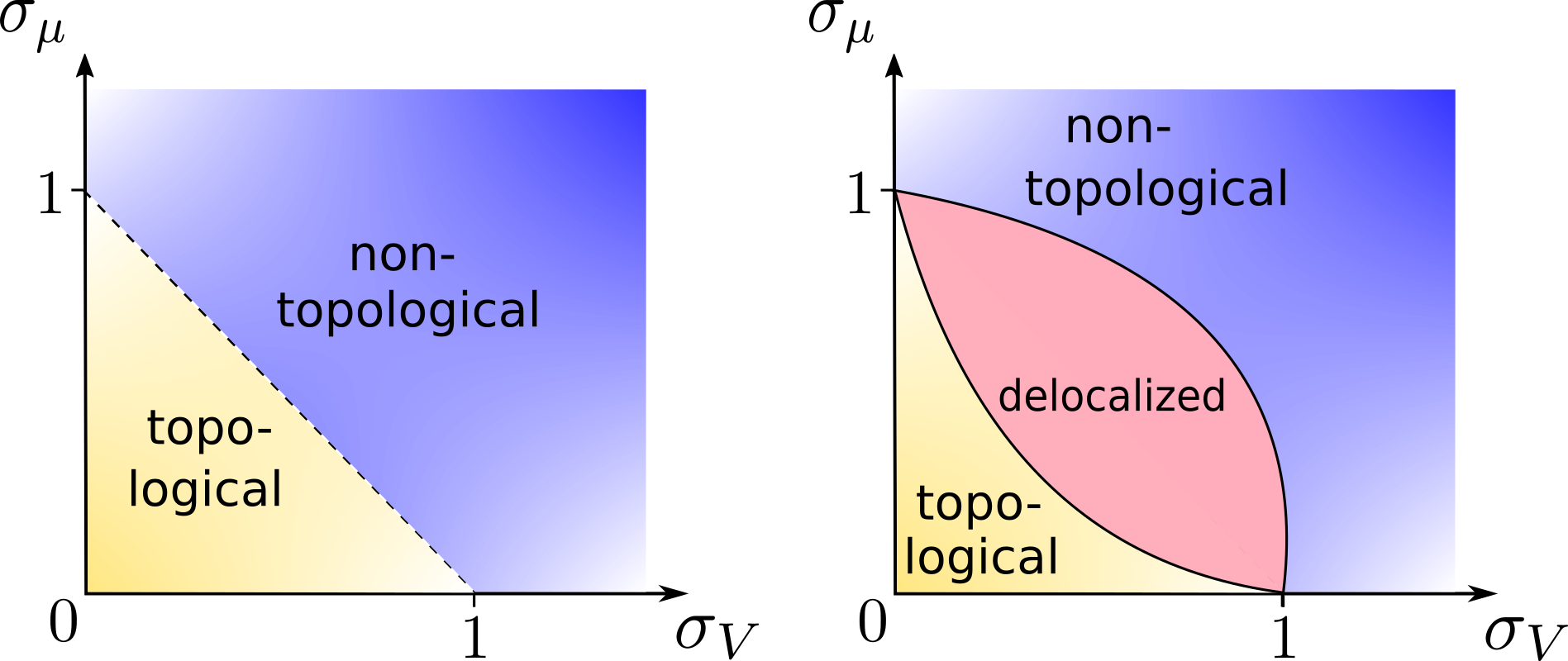}}
\put(-2,95){\textbf{b}}
\put(105,95){\textbf{c}}
\end{picture}
  \caption{a: Couplings in the Hamiltonian Eq.~\eqref{eq:Ham} with (t)LIOMs. 
  Black dots denote Majorana fermions and blue disks the stabilizers underlying the two types of LIOMs. 
  The stabilizers for conventional LIOMs involve Majorana pairs on physical sites; those for topological LIOMs have Majorana pairs straddling physical sites.
b, c: Possible phase diagrams. }    
\label{fig:phase_diagrams}
\end{figure}

The disordered interacting Kitaev chain~\cite{Kitaev_chain}  is a system of Majorana fermions $\gamma_n = \gamma_n^\dg$ with $\{\gamma_m,\gamma_n\} = 2\delta_{mn}$; 
the Hamiltonian is (cf. Fig.~\ref{fig:phase_diagrams}a)
\begin{align}
H &= \sum_{j=1}^{N-1} i t_j \gamma_{2j} \gamma_{2j+1} + \sum_{j=1}^N i \mu_j \gamma_{2j-1} \gamma_{2j} \notag \\*
&+ \sum_{j=1}^{N-1} V_j \gamma_{2j-1} \gamma_{2j} \gamma_{2j+1} \gamma_{2j+2}, \label{eq:Ham}
\end{align}
where the tunnel amplitude $t_j$, the on-site potential $\mu_j$ and the interaction strength $V_j$ are Gaussian distributed random variables with zero mean and standard deviation $\sigma_t = 1$, $\sigma_\mu$ and $\sigma_V$, respectively. 
The Hamiltonian commutes with the fermion parity operator $Z = \prod_{n = 1}^{2N} \gamma_n$, which splits the Hilbert space into two parity sectors. 

For $\sigma_\mu,\sigma_V \gg 1$,   the system is in the trivial phase with commuting projector representative
\begin{align}
H^\text{triv} &=  \sum_{j=1}^N i \mu_j \gamma_{2j-1} \gamma_{2j} +  \sum_{j=1}^{N-1} V_j \gamma_{2j-1} \gamma_{2j} \gamma_{2j+1} \gamma_{2j+2}.
\end{align}
$H^\mathrm{triv}$ may be expressed in terms of the stabilizers $S^\text{triv}_j = i \gamma_{2j-1} \gamma_{2j}$,  $j = 1, \ldots, N$; 
%
%it forms the commuting projector representative of the topologically trivial phase. 
%
consequently, its eigenstates correspond to the occupation of fermionic modes associated to physical sites, cf. Fig.~\ref{fig:phase_diagrams}a.

For $\sigma_\mu = \sigma_V = 0$ the system is in the topological phase. 
\begin{align}
    H^\text{topo} =\sum_{j=1}^{N-1} i t_j \gamma_{2j} \gamma_{2j+1}, 
\end{align}
with $S^\text{topo}_j = i \gamma_{2j} \gamma_{2j+1}$, $j = 1, \ldots, N - 1$,  
gives the commuting projector representative of this phase.
The eigenstates again correspond to the occupation of fermionic modes, but now they come from pairs of Majorana modes straddling physical sites, cf. Fig.~\ref{fig:phase_diagrams}a.
$S^\mathrm{nl} = i \gamma_{2N} \gamma_1$ forms a zero-energy fermion mode, resulting in a two-fold degenerate energy spectrum.

The phases are also known along the axes of the phase diagram: 
For $\sigma_V = 0$ the model is non-interacting and is in the topological (trivial) phase for $\sigma_\mu < 1$ ($\sigma_\mu > 1$)~\cite{Fisher92,Shankar87,Fisher95}. 
Similarly, for $\sigma_\mu = 0$, there is a phase transition point at $\sigma_V = 1$~\cite{Miao2017,McGinley2017}.
The two MBL phases can either be separated by a critical line or a delocalized phase~\cite{1DSPTMBL,Sze20}, cf. Fig.~\ref{fig:phase_diagrams}b,c.

\section{ED and order parameter benchmarks} 

\begin{figure}
\includegraphics[width=.48\textwidth]{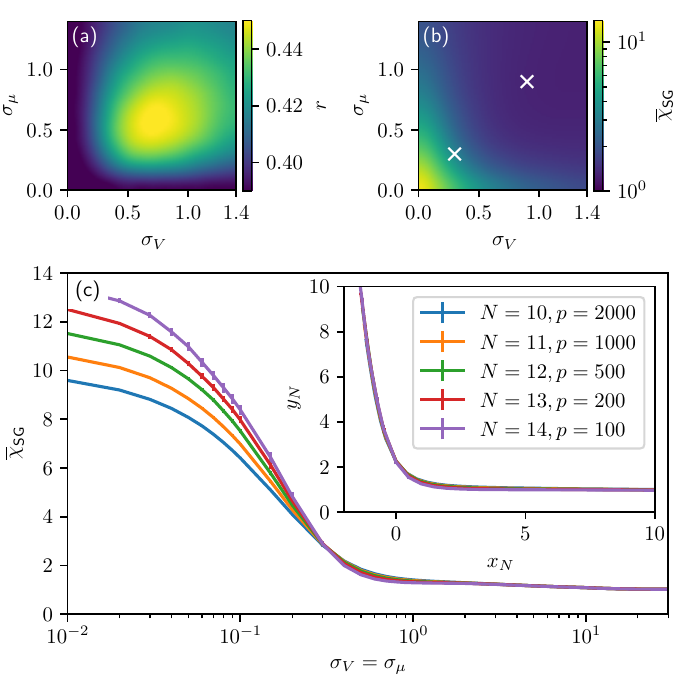}
  \caption{Exact diagonalization results.
  In all panels, we show averages over $p$ disorder realizations, both parity sectors, and the mid-third energies. 
  a: Gap ratio and b: average spin-glass order parameter $\overline \chi^\mr{SG}$, both for $p=100$ and system size $N = 14$.
  The maximal standard error of the mean (taken across disorder realizations in all panels) for any value of $\sigma_V$, $\sigma_{\mu}$ is 0.005 in (a) and 0.22 in (b). 
  c: $\overline \chi^\mr{SG}$ for different $N$ and $p$ along the diagonal $\sigma_V = \sigma_\mu$.
  The error bars mark the standard error of the mean.
  A scaling collapse~\cite{kjall2014many} with $x_N = (\sigma_\mu - \sigma_c) N^{0.6}$, $y_N = \overline \chi^\mr{SG}/N^{0.1}$ is shown in the inset. 
  The extracted critical point $\sigma_c = 0.3$ and the one for the dual order parameter, $\sigma_c' = 0.9$ (see Appendix~\ref{appB}), are marked by crosses in subfigure~b. 
}    
\label{fig:level_splitting}
\end{figure}

To provide a benchmark for our subsequent tLIOM analysis, we use ED to locate the MBL phases.  
We first analyze the level spacing for $N = 14$. 
In Fig.~\ref{fig:level_splitting}a we show the gap ratio $r_n = \min(s_{n-1},s_{n})/\max(s_{n-1},s_{n})$ with $s_n = E_{n} - E_{n-1}$ the level spacing \textit{in a given parity sector}. 
For each data point we average over the mid-third of energies (as those best reflect whether the system is thermalizing~\cite{Luitz2015,kjall2014many,Sze20}) and 100 disorder realizations. 
In an MBL phase we expect $r_\mathrm{P} = 0.386$~\cite{pal2010mb} (Poisson distribution) due to the lack of level repulsion; 
in a thermal phase we expect $r_\mathrm{WD} = 0.530$ (Wigner-Dyson distribution). 
Fig.~\ref{fig:level_splitting}a indicates that there are two MBL phases (with $r \approx r_P$), one for $\sigma_\mu + \sigma_V \lesssim 1$ and another one for $\sigma_\mu \gtrsim 1$ or $\sigma_V \gtrsim 1$.  
Although the gap ratio never gets as large as $r_\mathrm{WD}$, between these regions our results are consistent with an extended delocalized phase~\cite{Sze20,Moudgalya20,Sahay20}.

Topological properties can be detected by the spin-glass order parameter~\cite{Huse2013LPQO,Pekker2014HG,kjall2014many} $\chi_n^{\mr{SG}} = \frac{1}{N} \sum_{i,j=1}^N \langle n | \sigma_i^x \sigma_j^x | n \rangle^2$ in eigenstates $|n\rangle$ of the quantum Ising chain linked to our system via Jordan-Wigner transformation (cf. Appendix~\ref{appA}). 
While this is a two-point correlator of local operators in the spin language, it is a non-local order parameter (with $\gamma_j$ strings), as befits one detecting topological features, for the fermionic system.
We also use a ``dual order parameter" $\chi_n^\mr{dSG} = \frac{1}{N} \sum_{i,j=1}^N \langle n | \tilde \sigma_i^x \tilde \sigma_j^x | n \rangle^2$
where $\tilde \sigma_j^x$ are analogous to disorder operators~\cite{Fradkin78} in the Ising chain (cf. Appendix~\ref{appA}). 
$\chi_n^\mr{SG} \propto N$ for $N \gg 1$ in the topological phase, while $\chi_n^\mr{SG} \rightarrow 1$ in the trivial phase.
Conversely, $\chi_n^\mr{dSG} \propto N$ for $N \gg 1$ in the trivial phase, while $\chi_n^\mr{dSG} \rightarrow 1$ in the topological phase.
Our results for $\chi_n^\mr{(d)SG}$ (see Fig.~\ref{fig:level_splitting}b,c and Appendix~\ref{appB}) suggest that the $\sigma_\mu + \sigma_V \lesssim 1$ phase is topological,  while the other MBL phase is trivial. 

\section{Phase diagram from tLIOMs} 

\begin{figure}
\includegraphics[width=0.48\textwidth]{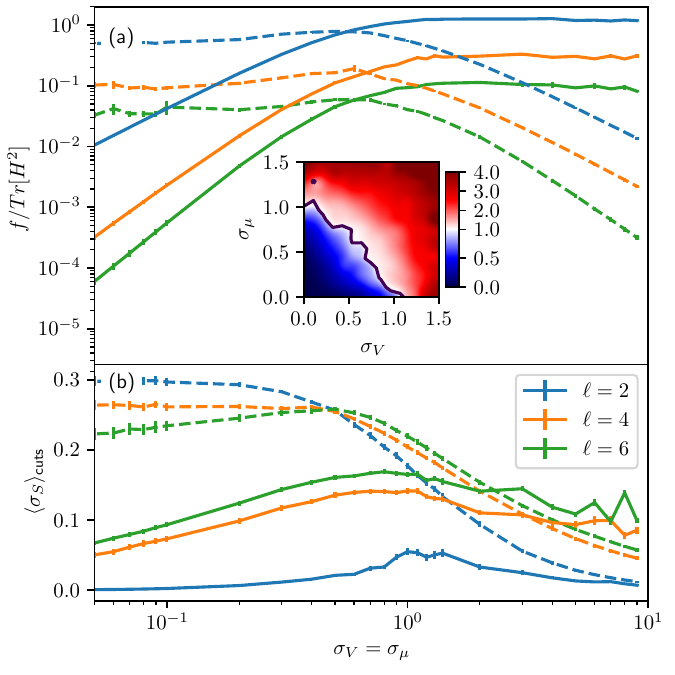}

  \caption{a: Figure of merit   $f/\tr(H^2)$ for topological (conventional) LIOMs, shown with solid (dotted) lines, for $N = 48$, averaged over 40 disorder realizations. 
  The error bars mark the standard error of the average across the different disorder realizations.
  The inset shows the ratio $f^{(t)}/f$ for $N = 48$, averaged over 20 disorder realizations and using $\ell = 6$. The contour ($f^{(t)} = f$)  separates the topological and trivial MBL phase, but is agnostic about the extent of the delocalized phase.  
b: Standard deviation $\sigma_S$ of the approximate-eigenstate-averaged entanglement entropy with respect to disorder realizations and subsequent averaging over entanglement cuts~\cite{Wahl2017PRX} using the same data as in subfigure a.
The error bars mark the standard error of $\langle \sigma_S \rangle_{\text{cuts}}$ across the cut positions.
  }    
\label{fig:ratio}
\end{figure}

We now apply the tLIOM approach to the model. 
We optimize, using the algorithm of Ref.~\onlinecite{Wahl2017PRX},  the quantum circuits $U_\mr{SR}$ and $U_\mr{SR}^{(t)}$  (for approximate LIOMs and tLIOMs, respectively) over the space of fermion parity conserving unitaries. 
We use two-layer quantum circuits with gates acting on $\ell$ sites each. 
In Fig.~\ref{fig:ratio}a, we show the resulting normalized figure of merit $f$ ($f^{(t)}$) for conventional (topological) LIOMs, for system size $N = 48$, gate lengths $\ell = 2,4,6$  and focusing on $\sigma_V = \sigma_\mu$. 
We also show  $f^{(t)}/f$ for $\ell = 6$ in the entire two-parameter phase diagram. 
As a function of $\ell$, we see a roughly exponential improvement deep in the MBL phases for the corresponding set of LIOMs.

Another way to map out the phase diagram using (t)LIOMs is entanglement entropy fluctuations~\cite{kjall2014many,Wahl2017PRX}.
We cut the approximate matrix product eigenstates, obtained from the optimized (t)LIOM ans\"atze, at a point at least ${3\ell}/{2}$ sites away from the boundary and apply the algorithm of Ref.~\onlinecite{Wahl2017PRX} to compute the average entanglement entropy. 
(Here, we averaged over all approximate eigenstates, as their entropies depend only on the expectation values of the tLIOMs near the cut~\cite{Wahl2017PRX}, which are not directly linked to overall energies.)
We repeat this step for all disorder realizations and calculate the standard deviation $\sigma_S$ of these averages. Finally, to reduce statistical fluctuations, we average $\sigma_S$ over all cut positions to obtain $\langle \sigma_S \rangle_\text{cuts}$.

We expect $\langle \sigma_S \rangle_\text{cuts}\rightarrow 0$  for $\sigma_V = \sigma_\mu \rightarrow \infty$, because there the LIOMs are simply the trivial stablizers (exact LIOM ansatz with $U_\text{SR} = \mathbb{1}$): the eigenstates are independently populated local fermion modes. 
Similarly, $\sigma_V = \sigma_\mu = 0$ is the topological MBL limit with tLIOMs simply the topological stabilizers (exact tLIOM ansatz with $U_\text{SR}^{(t)} = \mathbb{1}$),  independently of the disorder realizations, which again implies $\langle \sigma_S \rangle_\text{cuts} = 0$.
Away from these limits, but still in an MBL phase, i.e., with all eigenstates MBL, the entanglement and its  fluctuation are low due to the area law.  
In an ergodic phase, where all eigenstates are volume-law entangled, the equivalence of spectral and ensemble (i.e., over disorder) averages~\cite{BrodyRMT81} implies suppressed entanglement fluctuations.
In a thermal but not ergodic phase, with a $0<\nu<1$ fraction of volume-law states, we expect entanglement fluctuations to diverge in the thermodynamic limit due to $\nu$ being sensitive to small changes in the disorder~\cite{kjall2014many}.

The maximum amount of entanglement allowed by our approximation is proportional to $\ell$; hence, we expect $\langle \sigma_S \rangle_\text{cuts}$ to acquire a maximum, increasing with $\ell$~\cite{Wahl2017PRX}, where entanglement fluctuations would diverge. 
(This is analogous to the finite-size scaling in Ref.~\onlinecite{kjall2014many}.) 
This is consistent with Fig.~\ref{fig:ratio} if one combines the LIOM and tLIOM data in the regimes where they are reliable.

Based on the comparison between the optimized figures of merit for $\sigma_\mu = \sigma_V$, we expect the topological MBL phase to extend at most up to $\sigma_\mu = \sigma_V \lesssim 0.5$ and the trivial MBL phase to start above 
$\sigma_\mu = \sigma_V \gtrsim 0.5$. 
But the broad maxima of the entanglement entropy fluctuation of $U_\mr{SR}^{(t)}$ at $\sigma_\mu = \sigma_V \approx 0.5$ can indicate an interstitial thermal phase.\footnote{To definitively establish whether this interstitial thermal phase is present, one may seek clearer maxima using larger $\ell$. However, this is beyond the scope of this work; our focus is on detecting topological MBL.}
This is also consistent with conventional and topological LIOMs performing comparably in a window around $\sigma_\mu = \sigma_V \approx 0.5$. 
Outside of this window, however, the performance difference between conventional and topological LIOM ans\"atze allows us to clearly detect the topology of the MBL phase.

The picture obtained from our (t)LIOM approach is thus consistent with the (dual) spin-glass order parameter and the gap ratio from ED.
While meeting this benchmark, using (t)LIOMs we could probe system sizes well beyond the reach of ED. 
This suggests that our method can generalize well for detecting topological MBL phases beyond 1D.

\section{Topological doublets} 

Another topological MBL feature is the presence of topological multiplets across the entire energy spectrum~\cite{2013Bauer_Nayak,Huse2013LPQO}.
(The multiplet splitting is exponentially small in system size.)
In the Kitaev chain, one has topological doublets~\cite{Kitaev_chain,Huse2013LPQO}.
Here we show how these can be detected using tLIOMs.
A key ingredient is the access to the approximate nonlocal $\tau^\mathrm{nl} = U_\mr{SR}^{(t)} S^\mathrm{nl} U_\mr{SR}^{(t)\dg}$ and the corresponding flip operator (an MBL strong zero mode~\cite{Fendley}) $\tau^{\mathrm{nl},x} = U^{(t)}_\mathrm{SR} \gamma_1 U^{(t)\dagger}_\mathrm{SR}$: the approximate eigenstate $|\psi_\alpha\rangle$ forms a doublet with  $\tau^{\mathrm{nl},x} |\psi_\alpha\rangle$. 
The goal is to show that these two states have approximately the same energy. 
To enhance energy accuracy, and to get a method that can access spatial features beyond $\ell$ (which is required for studying multiplet splittings in large systems, see Appendix~\ref{appC}), we couple the tLIOM ansatz with DMRG-X~\cite{Khemani2016MPS}:  
we write $|\psi_\alpha\rangle$ and $\tau^{\mathrm{nl},x} |\psi_\alpha\rangle$ as matrix product states to use as inputs to DMRG-X. 
We denote the resulting outputs by $|\Psi_\alpha\rangle$ and $|\Psi^{\tau}_\alpha\rangle$, respectively. 
As DMRG-X is biased towards low-entanglement states~\cite{Devakul2017}, we must prevent convergence to $|\Psi^{(\tau)}_\alpha\rangle$ with different tLIOM expectation values than $|\psi_\alpha\rangle$ but with lower entanglement. 
To control for this, a convenient proxy is to require high $\sqrt[N]{|\langle \Psi_\alpha|\psi_\alpha\rangle|}$
and  
$\sqrt[N]{|\langle \Psi^{\tau}_\alpha|\tau^{\mathrm{nl},x}|\psi_\alpha \rangle|}$
which for tensor network states are broadly interpretable as geometric mean ``per-site overlaps." 

\begin{figure}
	\includegraphics[width=0.48\textwidth]{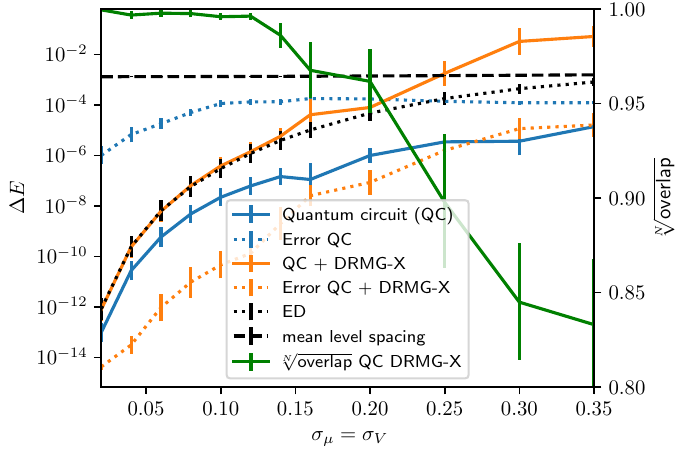}
	\caption{Doublet splittings from 
		DMRG-X-augmented tLIOMs, tLIOMs (QC), and ED for $N = 15$ and 10 disorder realizations. 
		The errors refer to the energy differences from the nearest ED energies.
		The tLIOM + DMRG-X results are averaged over 10 randomly chosen approximate eigenstates, while all other data are averaged over all eigenstates, for each disorder realization. 
		The green line shows the average per-site overlap between corresponding QC and DMRG-X-optimized states.
		Apart from the mean level spacing (dashed), all averages here use geometric means; this is due to splittings varying by orders of magnitude between realizations, and due to each per-site overlap being interpretable as a geometric mean. 
		The error bars mark the standard error of the geometric mean.
	}
	\label{fig:pairings}
\end{figure}

We now turn to demonstrating this approach. 
To be able to compare to ED, we work with $N = 15$ and deep in the topological MBL phase so that ED can assign almost all doublets correctly due to their small level splitting.
We optimize tLIOMs for 10 disorder realizations and for each realization we randomly choose 10 approximate eigenstates and their doublet partners.
In DMRG-X, we use bond dimensions up to 85. 
Our DMRG-X-augmented tLIOM results, direct tLIOM and ED data are shown in Fig.~\ref{fig:pairings}. 
The DMRG-X-augmented tLIOM splittings are in quantitative agreement with ED up to  $\sigma_V = \sigma_\mu\approx 0.13$.  
Beyond this, the breakdown of the method is signaled by the deteriorating per-site overlaps.  
(While we demonstrated its use for system sizes where a comparison to ED is available, the per-site overlap is expected to be a useful and feasibly computable indicator for system sizes beyond the reach of ED.)

Deep in the topological MBL phase, we expect this method to correctly predict the splittings for $N$ beyond the reach of ED.
The required bond dimensions will be larger, but still scale polynomially with $N$ due to having area-law eigenstates.
Beyond 1D, one could similarly take an approximate eigenstate obtained from the tLIOM approach~\cite{2DMBL,2DfMBL} in the topological phase as a starting point for an approximation using projected entangled pair states (PEPS)~\cite{PEPS} methods~\cite{Verstraete2008}. 
Although the mean level spacing is exponentially smaller than the multiplet splitting~\cite{Parameswaran2018}, the locality of PEPS will allow one to ensure convergence to the correct eigenstates.

\section{Conclusions}
 
We showed how tLIOMs can detect topological MBL in numerical simulations. 
Using (t)LIOM numerics for the interacting disordered Kitaev chain (up to $N = 48$), we found two topologically distinct MBL phases and features consistent with an intervening thermal phase. 
We also showed how to detect topological multiplets using tLIOMs which, for strong MBL and when augmented by DMRG-X, reached very high energy accuracy. % comparable to ED. 
It would be interesting to use tLIOMs above 1D (e.g., for the toric code) to detect MBL topological order and to detect topological multiplets, which here have splitting decaying slower with system size than the mean level spacing even for strong MBL.

\textit{Acknowledgments.--} 
We thank Nicolas Laflorencie for very helpful discussions. 
This work was supported by the EPSRC grant EP/S019324/1 and the ERC Starting Grant No.678795 TopInSy.
Our computations used resources provided by the Cambridge Service for Data Driven Discovery (CSD3) operated by the University of Cambridge Research Computing Service (\href{www.csd3.cam.ac.uk}{www.csd3.cam.ac.uk}), provided by Dell EMC and Intel using Tier-2 funding from the EPSRC (capital grant EP/P020259/1), and DiRAC funding from the Science and Technology Facilities Council (\href{www.dirac.ac.uk}{www.dirac.ac.uk}).

\appendix

\section{Ising chains via Jordan-Wigner transformations}\label{appA}

We rewrite the Hamiltonian $H$ using a Jordan-Wigner transformation, $\gamma_{2j-1} = (\prod_{k=1}^{j-1} \sigma^z_k) \sigma_j^x$, $\gamma_{2j} = -(\prod_{k=1}^{j-1} \sigma^z_k) \sigma_j^y$, $j = 1,2,\ldots,N$, 
\begin{align}
	H = \sum_{j=1}^{N-1} t_j \sigma_j^x \sigma_{j+1}^x + \sum_{j=1}^N \mu_j \sigma_j^z - \sum_{j=1}^{N-1} V_j \sigma_{j}^z \sigma_{j+1}^z, \label{eq:Ham_org}
\end{align}
with Pauli operator $\sigma^\alpha_i$, $\alpha = x,y,z$ acting on site $i$. Due to fermion parity conservation, the Hamiltonian is invariant under the parity operator $Z = \prod_{k=1}^N \sigma_k^z$. 
This generalized quantum Ising Hamiltonian is the same as the ``$hJJ'$ model'' studied in Ref.~\onlinecite{Pekker2014HG}. 
Its authors introduced a real-space renormalization group method for exited states, predicting two distinct MBL phases separated by a transition visible in the entire energy spectrum. 

An alternative Jordan-Wigner transformation $\gamma_{2j} = (\prod_{k=1}^{j-1} \tilde \sigma^z_k) \tilde \sigma_j^x$, $\gamma_{2j+1} = -(\prod_{k=1}^{j-1} \tilde \sigma^z_k) \tilde \sigma_j^y$ ($\gamma_1 = -(\prod_{k=1}^{N-1} \tilde \sigma_k^z) \tilde \sigma_N^y$) results in Kramers-Wannier-like duality, such that
\begin{align}
	H &= \sum_{j=1}^{N-1} t_j \tilde \sigma_j^z + \sum_{j=2}^{N} \mu_j \tilde \sigma_{j-1}^x \tilde \sigma_j^x - \sum_{j=2}^{N-1} V_j \tilde \sigma_{j-1}^x \tilde \sigma_{j+1}^x \notag \\
	&- \mu_1 \tilde Z \tilde \sigma_N^x \tilde \sigma_1^x + V_1 \tilde Z \tilde \sigma_N^x \tilde \sigma_2^x \label{eq:Ham_fid}
\end{align}
with $\tilde Z = \prod_{j=1}^N \tilde \sigma_j^z = -Z$ and $\tilde \sigma_j^\alpha$ likewise Pauli operators, but now acting on links. The two sets of Pauli operators are related via the transformation
\begin{align}
	\sigma_j^x &= \begin{cases} - \left(\prod_{k=1}^{N-1} \tilde \sigma_k^z\right) \tilde \sigma_N^y &\mbox{if } j < N, \\
		-\tilde \sigma_N^y & \mbox{if } j = N, \end{cases} \\
	\sigma_j^z &= \begin{cases} \tilde \sigma_{j-1}^x \tilde \sigma_{j}^x &\mbox{if } j > 1,\\
		-\tilde Z \tilde \sigma_1^x \tilde \sigma_N^x & \mbox{if } j = 1, \end{cases}
\end{align}
and inverse transformation
\begin{align}
	\tilde \sigma_j^x &= \begin{cases} -\sigma_1^y \left(\prod_{k=2}^{j} \sigma_k^z\right) &\mbox{if } j >1, \\
		-\sigma_1^y & \mbox{if } j = 1, \end{cases} \label{eq:inv1}\\
	\tilde \sigma_j^z &= \begin{cases} \sigma_{j}^x \sigma_{j+1}^x &\mbox{if } j < N,\\
		-Z \sigma_1^x \sigma_N^x & \mbox{if } j = N. \end{cases} \label{eq:inv2}
\end{align}
(An actual Kramers-Wannier transformation, where the $-\sigma_1^y$ in Eq.~\eqref{eq:inv1} are replaced by $\sigma_1^z$ and the second line in Eq.~\eqref{eq:inv2} by $\sigma_N^x$, changes only the last two terms of Eq.~\eqref{eq:Ham_fid}, which become local boundary terms.) 
In each symmetry sector of $\tilde Z$ (even / odd parity) Eq.~\eqref{eq:Ham_fid} is a generalization of the quantum Ising model with next-nearest neighbor coupling considered in Ref.~\onlinecite{kjall2014many} (after an on-site unitary transformation sending $\sigma_i^x \leftrightarrow \sigma_i^z$). 
There, the authors likewise found two distinct MBL phases as evidenced by spin-glass order. 
We note that for $\mu_j = 0$ $\forall \ j = 1,\ldots, N$ the Hamiltonian in  Eq.~\eqref{eq:Ham_fid} decouples \textit{in each parity sector} into two non-interacting chains, cf. the  structure of Eq.~\eqref{eq:Ham_org} for $\sigma_V = 0$~\cite{Miao2017,McGinley2017}. 
Hence, there is a phase transition point at $(\sigma_V,\sigma_\mu) = (1,0)$.

\begin{figure}
	\begin{picture}(240,160)
		\put(0,0){\includegraphics[width=0.46\textwidth]{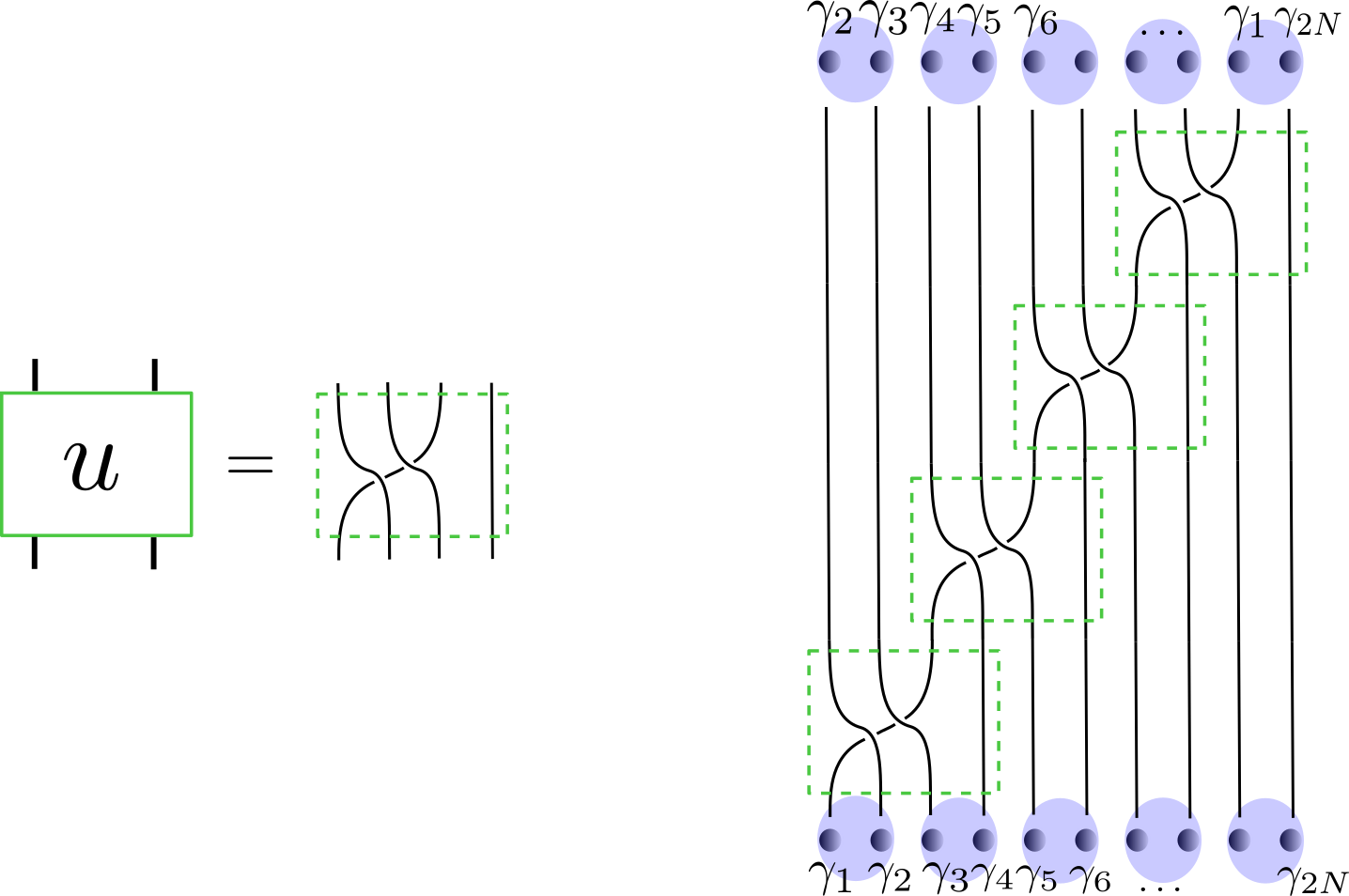}}
		\put(-2,150){\textbf{a}}
		\put(120,150){\textbf{b}}
	\end{picture}
	\caption{Unitary mapping between the bases $\{\sigma_i^\alpha\}$ and $\{\tilde \sigma_i^\alpha\}$. a: Constituting unitaries $u$ acting on 2 sites, i.e., 4 Majorana fermions. b: The tensor network of the overall unitary is constructed as a ``staircase'' of the small unitaries $u$ (denoted by dashed boxes), giving rise to the shown Majorana braid.}    
	\label{fig:permutation}
\end{figure}

The bases $\{\sigma_j^\alpha\}$ and $\{\tilde \sigma_j^\alpha\}$ are related via a deep quantum circuit, namely the one made up of a ``staircase'' of unitaries acting on nearest neighbors. 
In the fermionic picture these unitaries permute the Majorana fermions as $(\gamma_1, \gamma_2, \gamma_3, \gamma_4) \rightarrow (\gamma_2,\gamma_3,\gamma_1,\gamma_4)$, see Fig.~\ref{fig:permutation}. 

\section{Order parameters}\label{appB}

\begin{figure}
	\includegraphics[width=.48\textwidth]{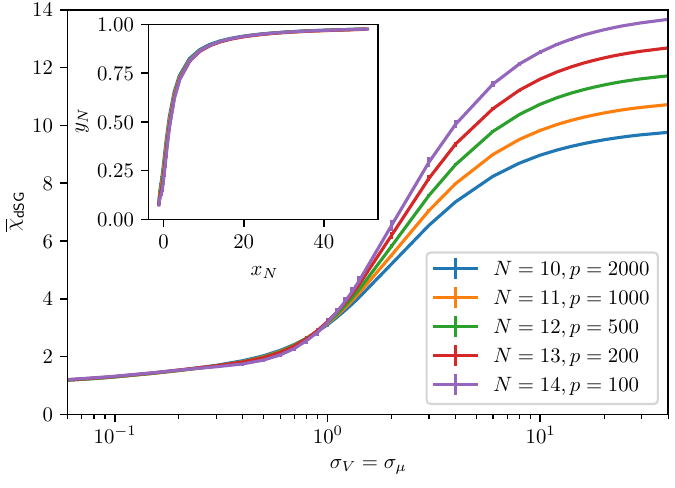}
	\caption{ED results for $\overline \chi^\mr{dSG}$ for different system sizes $N$ averaged over $p$ disorder realizations along the diagonal $\sigma_V = \sigma_\mu$. 
		We show a scaling collapse with $x_N = (\sigma_\mu - \sigma_c') N^{0.1}$ and $y_N = \overline \chi^\mr{dSG}/N^{1.0}$ in the inset. 
		The extracted critical point $\sigma_c' = 0.9$ is marked by a cross in Fig.~2b. 
		The error bars mark the standard error of the mean.
	}    
	\label{fig:dual}
\end{figure}

The spin-glass order parameter~\cite{Huse2013LPQO,Pekker2014HG,kjall2014many} measures the presence of regions with fixed $\sigma^x$-magnetization separated by domain walls and thus detects the spin-glass phase. It is given by
\begin{equation}
	\chi_n^\mr{SG} = \frac{1}{N} \sum_{j,k=1}^N \langle n | \sigma_j^x  \sigma_k^x | n \rangle^2
\end{equation}
for the eigenstate $|n\rangle$. 
In terms of Majorana fermions, 
\begin{equation}
	\chi_n^\mr{SG}= 1 + \frac{2}{N} \sum_{k>j}^N (-1)^{k-j} \langle n| \prod_{i=j}^{k-1}\gamma_{2i} \gamma_{2i+1}|n\rangle^2
\end{equation}
involves topological Majorana strings, i.e., contiguous products of topological stabilizers $S_j=i\gamma_{2j}\gamma_{2j+1}$ (cf. Fig.~1a of the main text).
The dual order parameter 
\begin{equation}
	\chi_n^\mr{dSG} = \frac{1}{N} \sum_{j,k=1}^N \langle n | \tilde \sigma_j^x \tilde \sigma_k^x | n \rangle^2
\end{equation}
detects spin-glass order in terms of the dual variables $\tilde \sigma_i^x$ in Eq.~\eqref{eq:Ham_fid}. 
In terms of Majorana fermions
\begin{equation}
	\chi_n^\mr{dSG}= 1
	+ \frac{2}{N} \sum_{k>j}^N (-1)^{k-j} \langle n| \prod_{i=j+1}^{k}\gamma_{2i-1} \gamma_{2i}|n\rangle^2\label{eq:chidual}
\end{equation}
involves strings of trivial stabilizers $S_j=i\gamma_{2j-1}\gamma_{2j}$ (cf. Fig.~1a of the main text).
Since the bases $\{\sigma_j^\alpha\}$ and $\{\tilde \sigma_j^\alpha\}$ are related by a unitary transformation exchanging stabilizer sets of complementary topological character (see Fig.~\ref{fig:permutation}), while $\chi_n^\mr{SG} \propto N$ (for $N\gg 1$) for MBL with topological order (and $\chi_n^\mr{SG} \rightarrow 1$ otherwise), we have $\chi_n^\mr{dSG} \propto N$ (for $N\gg 1$) for topologically trivial MBL (and $\chi_n^\mr{dSG} \rightarrow 1$ otherwise). This behavior can be seen in Fig.~\ref{fig:dual}, where we plot the dual order parameter as a function of $\sigma_\mu = \sigma_V$.

\section{Quantum circuit splitting for large systems}\label{appC}
In this section we demonstrate that the splittings corresponding to the quantum circuit ansatz are zero for sufficiently large systems; however, we note that the QC states are a good starting points for the DMRG-X calculation regardless.

The eigenstates that the quantum circuit approximation of tLIOMs predicts are given by
\begin{equation}
	\ket{\psi_\pm} = U_{\text{SR}} \ket{\bf{t}_\pm},
\end{equation}
where $\ket{\bf{t}_\pm}$ is an eigenstate of the topological stabilizers $S^\text{topo}_j = i \gamma_{2j} \gamma_{2j + 1}$ with eigenvalues $t_j = \pm 1$ and a $\pm 1$ eigenstate of the bi-local stabilizer $S^\mathrm{nl} = i \gamma_{2N} \gamma_1$. It may be written in terms of stabilizers as
\begin{equation}
	{\ket{\bf{t}_\pm}\bra{\bf{t}_\pm}} = \left(\frac{1}{2} \pm \frac{1}{2} S^\mathrm{nl}\right) \prod^{N - 1}_{j = 1}{\left(\frac{1}{2} + \frac{1}{2} t_j S^\text{topo}_j\right)}.
\end{equation}
Using this representation, the expectation value for the energy can be expressed as
\begin{align}
	&\bra{\psi_\pm} H \ket{\psi_\pm} \notag \\
	&= \Tr \left[ \left(\frac{1}{2} \pm \frac{1}{2} S^\mathrm{nl}\right) \prod^{N - 1}_{j = 1}{\left(\frac{1}{2} + \frac{1}{2} t_j S^\text{topo}_j\right)}U_{\text{SR}}^\dagger H U_{\text{SR}} \right]
\end{align}
and the corresponding splitting is given by
\begin{align}
	&\Delta E =  \bra{\psi_+} H \ket{\psi_+} - \bra{\psi_-} H \ket{\psi_-} \notag \\
	&= \Tr \left[ S^\mathrm{nl} \prod^{N - 1}_{j = 1}{\left(\frac{1}{2} + \frac{1}{2} t_j S^\text{topo}_j\right)}U_{\text{SR}}^\dagger H U_{\text{SR}} \right].
\end{align}
To further analyze the expression, we can expand the Hamiltonian in local terms
\begin{equation}
	H = \sum_{k=1}^N{h_k};
\end{equation}
with that the splitting is given by
\begin{equation}
	\Delta E = \sum_{k=1}^N \Tr \left[ S^\mathrm{nl} \prod^{N - 1}_{j = 1}{\left(\frac{1}{2} + \frac{1}{2} t_j S^\text{topo}_j\right)}U_{\text{SR}}^\dagger h_k U_{\text{SR}} \right].
\end{equation}
Since the quantum circuit acts locally, the causal cones \cite{Wahl2017PRX} of each term $h_k$ in the Hamiltonian will, for a sufficiently large system size $N$, not cover both of the Majorana modes making up $S^\text{nl}$. Hence, either $\gamma_1$ or $\gamma_N$ (or both) will factor from the trace, leading to $\Delta E = 0$.

\begin{figure}
	\includegraphics[width=.48\textwidth]{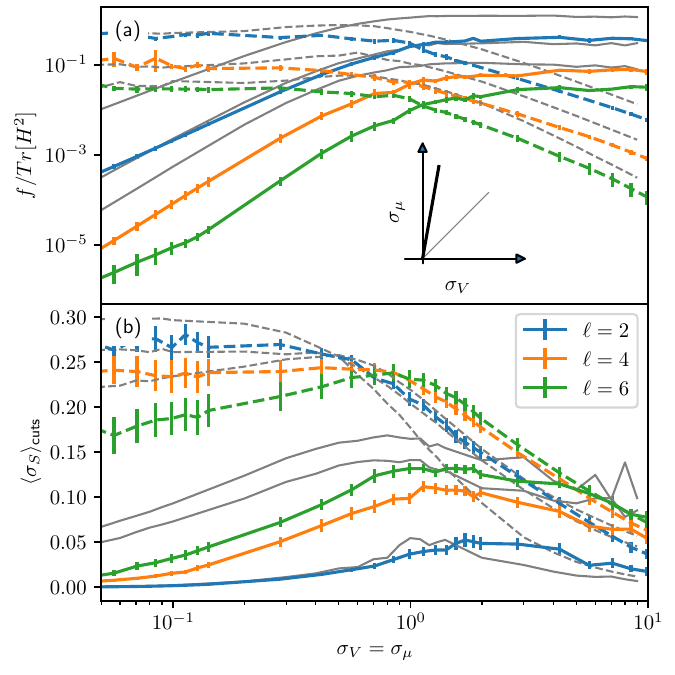}
	\caption{a: Figure of merit $f / \tr(H^2)$ and b: standard deviation of the entanglement entropy for topological  (conventional) LIOMs, shown with solid (dotted) lines, for 10 disorder samples and $N = 48$ evaluated along a line away from the diagonal in the phase diagram as indicated in the inset. For comparison we added the data from Fig.~3 in gray. In (a) the error bars show the standard error of the average across the different disorder realizations and in (b) the error bars show the standard error of $\langle\sigma_S\rangle_\text{cuts}$ across the cut positions.
	}    
	\label{fig:angle}
\end{figure}

The true splitting is of the order $\exp(-N/\xi_L)$, i.e., it is due to the exponentially decaying coupling between the end-modes in the bi-local integral of motion $\tau^\text{nl}$. It is thus not surprising that the QC approximation alone, which uses strictly short-range gates, cannot reproduce the splitting. We can capture $\Delta E$ better once the accuracy of the approximation has been  improved, e.g., by feeding the QC approximate eigenstates into the DMRG-X algorithm. As pairs of QC initial states have the same configuration of approximate tLIOMs (other than for $\tau^\text{nl}$) and the DMRG-X algorithm tries to maintain a high overlap with the initial states, it will build up additional entanglement around these and thus improve their accuracy. This keeps the configuration of tLIOMs fixed for sufficiently large bond dimensions. Since the DMRG-X algorithm is based on matrix product states, it is able to build entanglement across the entire chain and will thus reproduce the energy splittings, including their dependence on system size, even for large systems.

\section{Sensitivity to the thermal phase}\label{appD}
To further illustrate the predictive power of the tLIOM approach, we have evaluated the (t)LIOMs along the $\sigma_V / \sigma_{\mu} = \tan(10^\circ)$ line, where we expect a smaller extent of the thermal phase (cf. Fig~1c).
Along this line the figure of merit (Fig.~\ref{fig:angle}a) has a smaller transition region in which both the LIOMs and tLIOMs poorly describe the system, consistent with a larger regime being a part of an MBL phase. Furthermore, focussing on the entanglement entropy fluctuations (Fig.~\ref{fig:angle}b), we find that the maxima they assume are smaller. Since the thermal phase is responsible for generating those maxima, this is also consistent with a smaller extent.

\end{document}